\documentclass[12pt,a4paper]{article}
\usepackage{graphicx}
\usepackage{lineno}
\usepackage{amsmath}
\usepackage{amssymb}
\usepackage{bm}
\usepackage{natbib}
\usepackage{booktabs}
\usepackage{geometry}
\usepackage{graphicx}
\usepackage{todonotes}

\newcommand{\beq}{\begin{equation}}
\newcommand{\eeq}{\end{equation}}

\begin{document}

\title{\textsc{Numerical Representations of Metabolic Systems}}

\author{Age K. Smilde*$^{1}$ and Thomas Hankemeier $^2$}

\maketitle

\begin{tabular}{l}
$^1$Biosystems Data Analysis, Swammerdam Institute for Life Sciences, \\
University of Amsterdam, The Netherlands.\\
$^2$Analytical Biosciences, LACDR, Leiden University, The Netherlands.\\

$^*$Corresponding author (a.k.smilde@uva.nl)\\
\end{tabular}

\section{Abstract}
Metabolomics is becoming a mature part of analytical chemistry as evidenced by the growing number of publications and attendees of international conferences dedicated to this topic. Yet, a systematic treatment of the fundamental structure and properties of metabolomics data is lagging behind. We want to fill this gap by introducing two fundamental theories concerning metabolomics data: data theory and measurement theory. Our approach is to ask simple questions, the answers of which require applying these theories to metabolomics. We show that we can distinguish at least four different levels of metabolomics data with different properties and warn against confusing data with numbers.\\

\noindent Keywords: data theory, measurement theory, comparability, measurement scales.

\section{Introduction}
Metabolomics concerns the measurement of small biochemical compounds (metabolites) in samples obtained from biological systems, or, in a broader context, from samples that contain such metabolites (extracts from natural foods, environmental samples etc). Such measurements are subsequently used to infer relevant information about the associated (biological) system related to a certain research question.\\

\noindent Nowadays, there is a whole variety of metabolomics measurements available which can be categorized either by the type of instruments used (mostly LC-MS, GC-MS, CE-MS and NMR\footnote{For a list of abbreviations, see Table \ref{Table2} in Section \ref{List of abbreviations}}) or by the type of measurement performed. The latter pertains to whether the measurement is targeted to a certain number of (known) metabolites or to an untargeted analysis in which also (many) unknown metabolites are being measured. There are also methods which are a combination of both. A typical pipeline for a metabolomics study runs through different steps: formulating a biological question, experimental design, sampling, measuring, preprocessing the data, analyzing the preprocessed data, visualization of results and answering the biological question \citep{Koek2011}.\\

\noindent An often neglected part of the above mentioned pipeline is the difference between numbers and data. This is a very fundamental issue at the heart of any measurement. We will explain this issue starting from asking a few very simple questions about whether numbers in one metabolomics experiment can be meaningfully compared to each other. To answer these questions, we have to introduce two theories, namely data theory and measurement theory. After that, we will give (partly) answers to the questions and try to come to a synthesis. As a running example throughout this paper we will consider measuring lipids in blood using LC-MS.\\

\section{Simple questions}
\label{SimpleQuestions}
\begin{figure}[h]
 \includegraphics*[width=10cm]{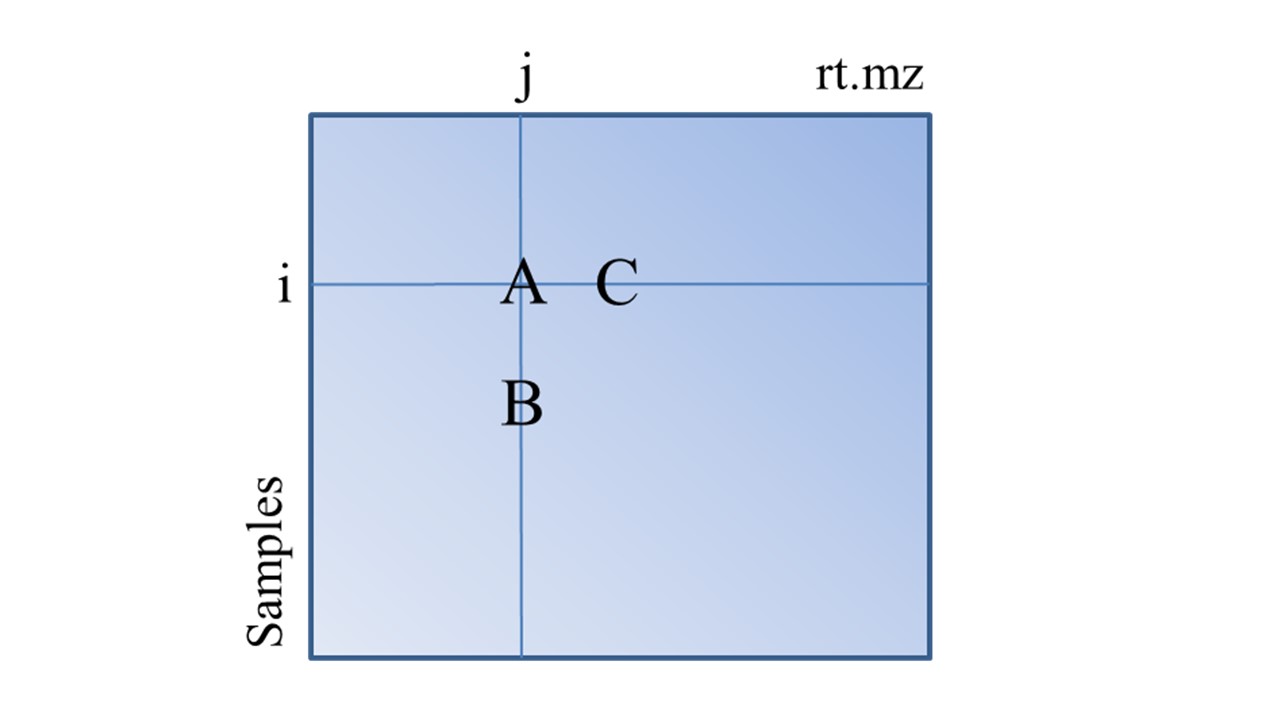}
 \caption{\footnotesize Schematic of raw measurements of lipids in blood. Legend: $i$ is a row in the matrix; $j$ is a column; $A$, $B$ and $C$ are specific numbers in the matrix; rt.mz is the retention time-mass spectrometry index.}
 \label{Figure1}
\end{figure}

We start by visualizing the numbers obtained from an LC-MS experiment on lipids in blood (see Figure \ref{Figure1}). The raw data can be arranged in intensities obtained at a certain m/z value at a certain retention time (rt) and the combined index rt.mz indicates a column in the matrix containing the samples in its rows. Looking at Figure \ref{Figure1}, we can ask the simple question to what extent the numbers are comparable, specifically:
\begin{enumerate}
  \item
    \begin{enumerate}
      \item If $A>C$; does that have a meaning?
      \item If $A>B$; does that have a meaning?
    \end{enumerate}
  \item
    \begin{enumerate}
      \item Does $A-C$ have a meaning?
      \item Does $A-B$ have a meaning?
    \end{enumerate}
  \item
    \begin{enumerate}
      \item Does $A/C$ have a meaning?
      \item Does $A/B$ have a meaning?
    \end{enumerate}
\end{enumerate}
which should be taken as examples, e.g., when $A<C$ then question one has to change accordingly. Note that moving from question one to three puts a higher demand on the numbers, e.g., if $A/C$ then necessarily $A>C$ must have a meaning (but not vice versa!). This notion will be formalized later.\\

\noindent The questions asked above are relevant for a subsequent data analysis. Take the example of PCA: the workhorse of metabolomics data analysis. The score-plots of a PCA are usually interpreted in terms of distances between dots representing the samples where samples far apart are regarded as very dissimilar and vice versa. But scores are linear combinations of the original variables (i.e. numbers) and this assumes that for distances in scores plots to be meaningful, at least also the original numbers should be comparable (at least at the level of $A-C$, to be explained later). Similar reasonings hold for loading plots, and for the results of PLS-DA and other often used techniques. Hence, it makes sense to answer the above posed questions.\\

\noindent Actually, there is even a more basic question to ask before considering the simple questions: is it even meaningful to start comparing the values A, B and C? This question is key in the field known as \emph{data theory}. The next questions regarding at which level comparisons are possible is the subject of \emph{measurement theory}. Therefore, both of these will be explained briefly in the sequel. This paper will be mainly concerned with mass-spectrometry based measurements; the case for NMR is a little different and will be touched upon in the end.

\section{Data theory}
A set of notions regarding comparability is called data theory and was pioneered by Coombs \citep{Coombs1964}and explained for multiway analysis \citep{VanMechelen2011}. The first important notion in data theory is conditionality, where we can distinguish column-, row-, and matrix-conditionality.\\

\begin{figure}[h]
 \includegraphics*[width=10cm]{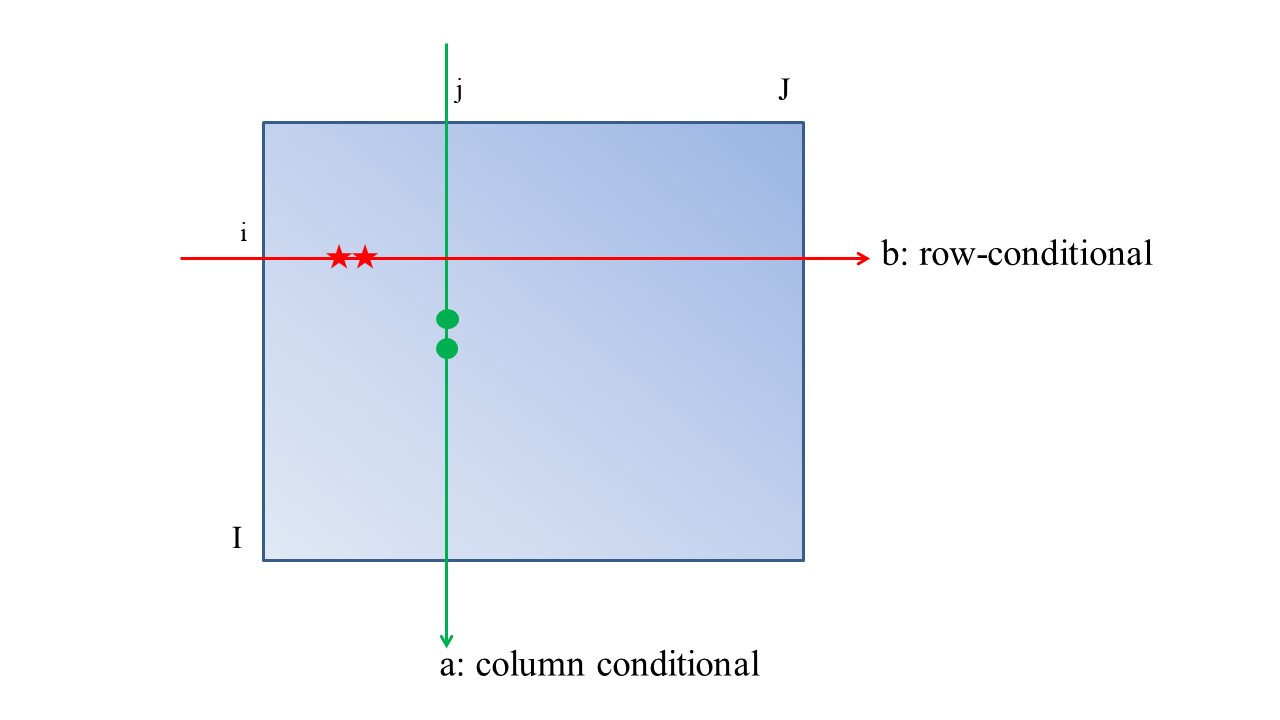}
 \caption{\footnotesize Column- and row-conditionality. Column-conditionality means that numbers in the same column can be compared (a); row-conditionality means that numbers in the same row can be compared (b).}
 \label{Figure2}
\end{figure}

\noindent When considering numbers arranged in a matrix (such as in Figure \ref{Figure1}) then different types of comparisons can be made: between numbers across rows in the same column (direction a) of Figure \ref{Figure2}) and between numbers across columns in the same row (direction b) of Figure \ref{Figure2}). When such data can be compared meaningfully, the data are called column-conditional and row-conditional, respectively. When data can be meaningfully compared across rows and columns, then these data are called matrix-conditional.\\

\noindent The prototypical example of row-conditional data are metabolomics measurements of urine, e.g., using NMR. Depending on the different urine histories of the subjects, the urine can be more or less concentrated. This makes the values within one column of a data matrix incomparable since the (unknown) dilution factor of the subjects destroys the comparability. The typical solution of this problem is found in normalizing the different samples thereby attempting to achieve matrix-conditionality. Whether this completely solves the problem is a matter of debate and it also depends on the research question. Actually, different types of metabolites are differently excreted by the kidney: some are only excreted by filtration, some are (partly) re-adsorbed, and re-adsorption is achieved by different transporters, for example one for acidic amino acids, one for dibasic amino acids, one for neutral amino acids \citep{Zelikovic1989}. This could justify a normalization per certain metabolite classes rather than normalizing all metabolites in the same manner.\\

\begin{figure}[h]
 \includegraphics*[width=10cm]{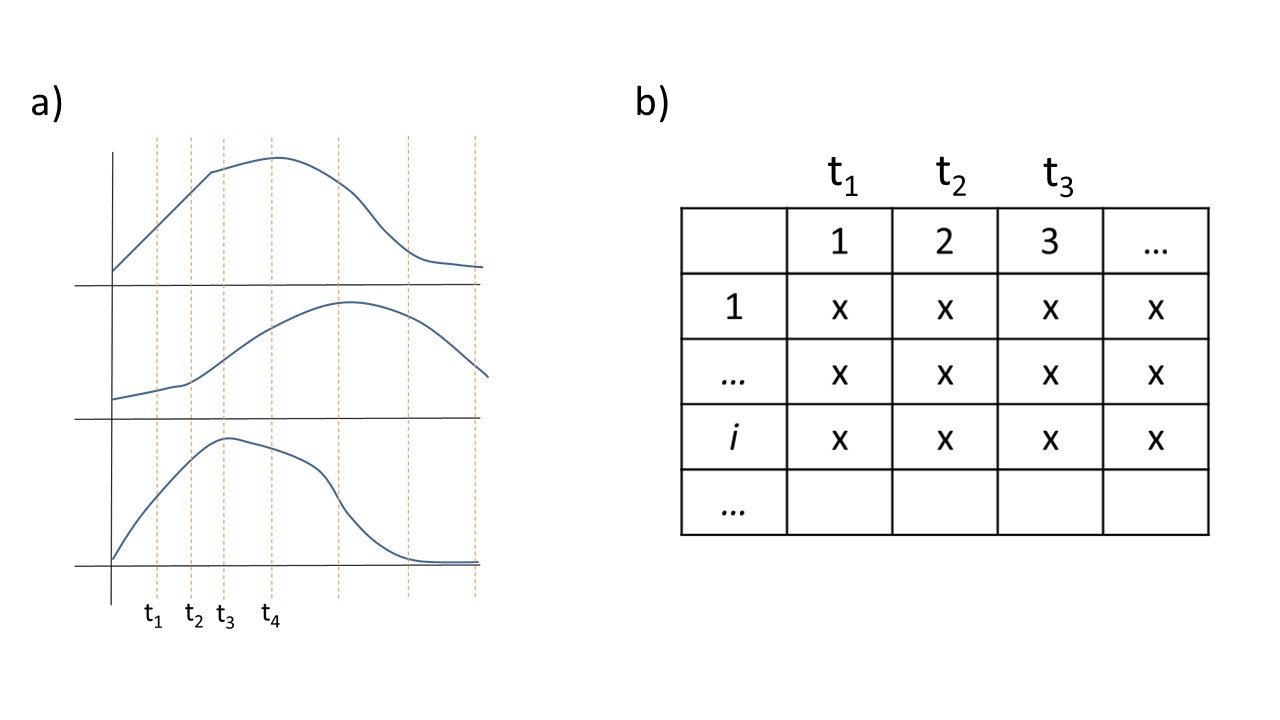}
 \caption{\footnotesize Lack-of-invariance illustrated. Panel a): unsynchronized times series of several subjects, b) the naive arrangement of the numbers.}
 \label{Figure3}
\end{figure}

\noindent A more serious problem regarding comparability as discussed in data theory is lack-of-invariance: the numbers in a single column do not have the same meaning. This problem is more fundamental than conditionality. Whereas in conditionality, numbers cannot be compared since there are (unknown) arbitrary differences, in lack-of-invariance the meaning of the variable changes within a column. The prototypical example is unsynchronized time series data (see Figure \ref{Figure3}, panel a)). Time series of three subjects are collected for multiple metabolites; in this figure only one metabolite is shown.  The series are not synchronized, therefore the measurements at, e.g., physical time point $t_4$ cannot be compared across subjects because they pertain to different states of the biological process measured with the metabolite. Hence, the meaning of the measurement at time point $t_4$ changes and is not invariant.\\

\begin{figure}[h]
 \includegraphics*[width=8cm]{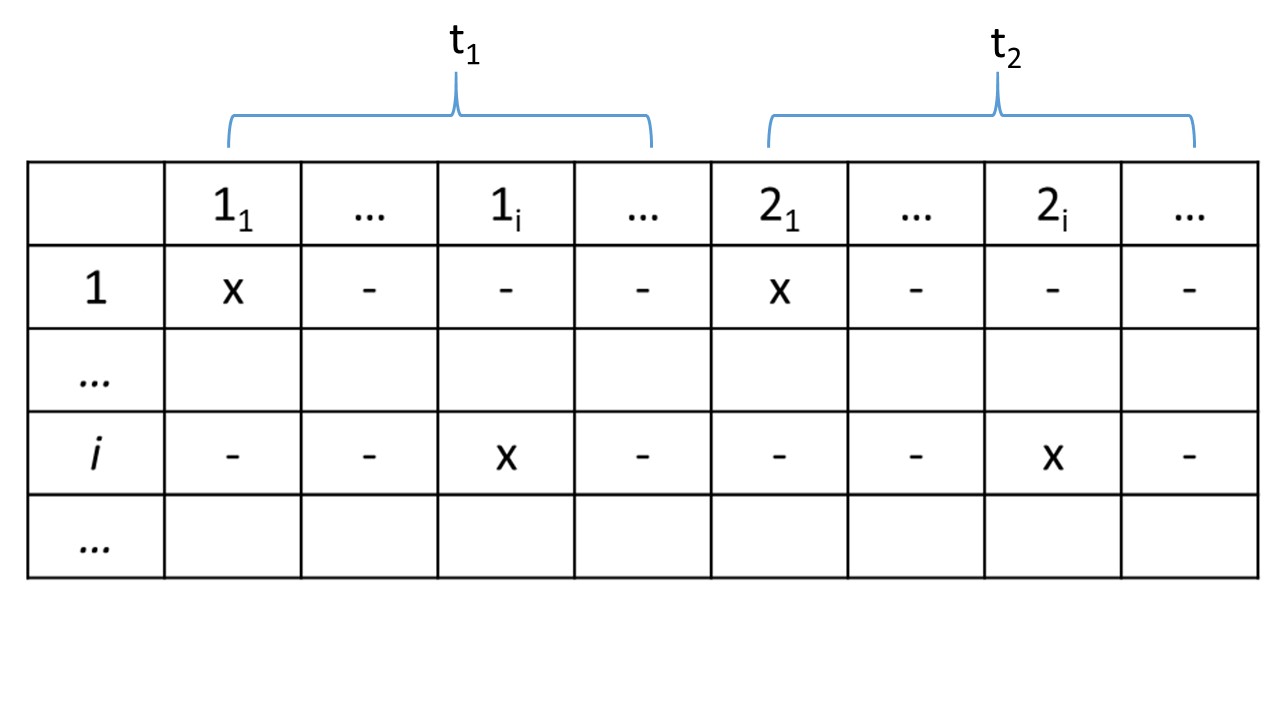}
 \caption{\footnotesize Proper arrangement of the numbers whereby each individual i receives its own time points.}
 \label{Figure4}
\end{figure}

\noindent A naive arrangement of the numbers is shown in Figure \ref{Figure3}, panel b). This is called naive since the lack-of-invariance is not taken into account. A more accurate arrangement of the numbers is shown in Figure \ref{Figure4} because now it is clear that each subjects has its own unique time points (see the subscript $i$ on the variables indicating time points). Obviously, the numbers as shown in the latter panel cannot be used as such. Remedies of this problem are found in alignment procedures (e.g. using warping approaches \citep{Christin2010}). After such an alignment of all metabolites - assuming that this has solved the lack-of-invariance problem -  the numbers can be arranged in a three-way array and analyzed with proper three-way methods \citep{Smilde2004}.\\

\noindent It is important to realize that such lack-of-invariance problems can also be solved by using data analysis methods that do not require the numbers to be synchronized. One such an alternative for the case of Figure \ref{Figure3}, panel a) is to concatenate the data sets per subject (time versus metabolites) in such a way that all subject-matrices are stacked on top of each other with the metabolites as the common mode. Then methods like Simultaneous Component Analysis \citep{Timmerman2003} can be used. This is one of the ways to solve the synchronization problems in batch statistical process monitoring where the batches are also not synchronized \citep{Wold1998}. Note that such an alternative arrangement of the numbers will have repercussions on the types of answers to be obtained from these numbers.

\section{Measurement theory}
After having established that a comparison between numbers is meaningful, the next question is at what level this can be done. This was pioneered by Stevens \citep{Stevens1946} and later taken up and further developed by several authors \citep{Krantz1971,Roberts1985,Narens1986,Luce1987}. One of the dominant theories of measurement is representational theory \citep{Hand1996} and this will be explained briefly in the following. There are two notions important in this theory: a \emph{representation} of a system and \emph{uniqueness} properties of the numerical representation of that system. A small example will be used to explain these ideas.\\

\begin{figure}[h]
 \includegraphics*[width=10cm]{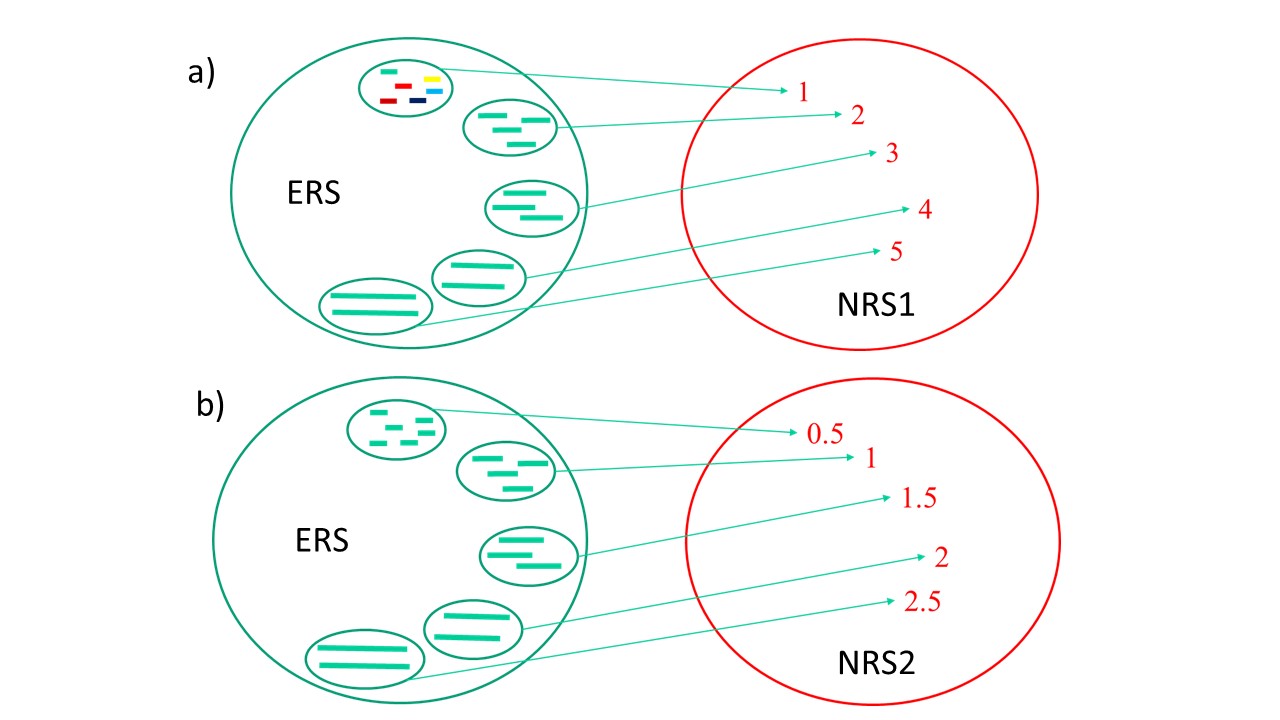}
 \caption{\footnotesize Numerical representations of the lengths of sticks: a) left: the empirical relational system (ERS) of which only the length is studied, right: a numerical representation (NRS1), b) an alternative numerical representation (NRS2) of the same ERS carrying essentially the same information.}
 \label{Figure5}
\end{figure}

\noindent Suppose we consider all sticks in the world; these are shown as an Empirical Relational System (ERS) in the upper left of Figure \ref{Figure5}. Although the sticks may have different colors, we are only interested in their lengths. The relationships between the sticks can be represented numerically with the numbers in the upper right panel of Figure \ref{Figure5}. An equally valid representation of the lengths of the sticks is given in the below right panel of Figure \ref{Figure5}. Hence, we have two numerical representational systems (NRS1 and NRS2) that can both represent the ERS. Although the two NRS's are different the ratio's between the numbers within both systems is the same: that property of the NRS is unique. Such systems (and associated measurements) are therefore called ratio-scaled measurements (the unit is arbitrary, see also Simple Question three). Likewise, it is possible to define interval-scaled measurements (e.g. for measuring temperatures of objects using degrees Celsius; unit and zero-point are arbitrary; see also Simple Question two) and ordinal-scaled measurements (e.g. degrees of agreement with statements; see also Simple Question one). At the lowest level, numbers are nominal and only indicate discrete categories. A more formal treatment of these concepts is given in the Section \ref{Formal treatment of measurement scales}.\\

\noindent When considering the simple questions, it is clear that metabolomics measurements can have different levels of measurement scales. It is certainly not always the case that metabolomics measurements are measured on a ratio-scale. This will be explained in Section \ref{Levels}. Two important types of measurements in chemistry (and, thus, in metabolomics) are concentrations and pH. Concentrations have special characteristics (see Section \ref{Concentrations}); and pH cannot be captured in representational theory but is defined in the context of operational theory (see Section \ref{pH}).

\section{Levels of metabolomics measurements}
\label{Levels}
\subsection{Level 0 measurements: raw numbers}
\label{Level 0 measurements}
\begin{figure}[h]
 \includegraphics*[width=10cm]{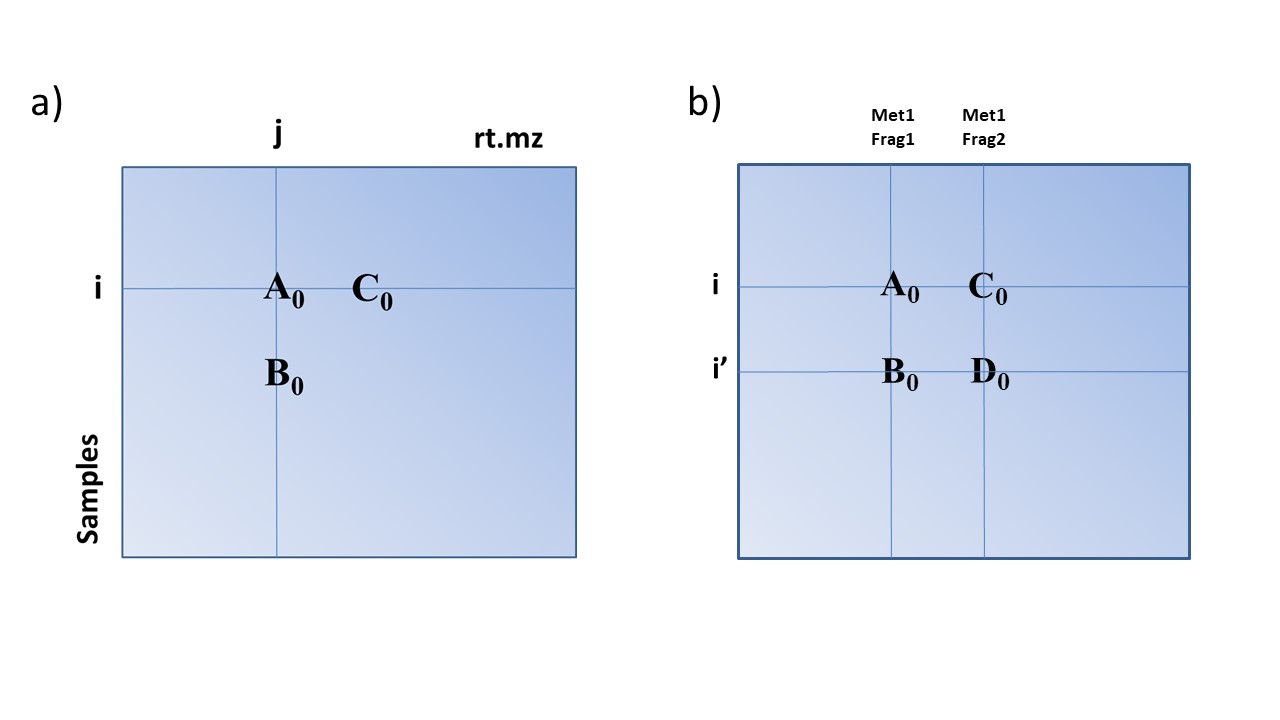}
 \caption{\footnotesize Level 0 of LC-MS measurements; a) rt.mz is a specific combination of retention time and m/z ratio; $i$ is an index for sample, $j$ is an index for column, b) Met1 means metabolite 1, Frag1 is fragment 1 and Frag2 is fragment 2 of the same metabolite 1.}
 \label{Figure6}
\end{figure}

Given the knowledge explained above regarding different aspects of comparability we now turn to the Simple Questions of Section \ref{SimpleQuestions}. The most basic measurement readouts of an LC-MS measurement of blood-lipids are shown in Figure \ref{Figure6}a. This is simply a list of raw intensities measured per sample in an LC-MS run arranged in an rt.mz format and will be called Level 0 numbers.\\

\noindent We can now start by answering the first Simple Question. Suppose that $A_0>C_0$: does that have a meaning? There are two cases to consider. Case a) where the numbers pertain to fragments of different metabolites (we do not consider trivial cases of spurious signals due to noise). For this case, the answer is that $A_0>C_0$ has no meaning since the response factors of both metabolites are different and at this point unknown. Hence, these numbers do not reflect (relative) concentrations within the system. Case b) is shown in Figure \ref{Figure6}b and pertains to intensities of different fragments\footnote{This does not hold for adducts; there ratio's can depend on the concentrations} of the same metabolite (and, thus, at the same rt). In that case, the ratio $A_0/C_0$ may have meaning since it refers to the same metabolite. In fact, such a ratio should also hold for the same fragments of that metabolite in other rows, thus $A_0/C_0=B_0/D_0$ (assuming alignment of rt.mz values). Comparing $A_0$ with $B_0$ in case a) we run into a lack-of-invariance problem since the rt.mz values are not aligned. Even when alignment would not be a problem, we still have only row-conditional numbers since there may be batch and sample work-up differences between the samples.\\

\begin{figure}[h]
 \includegraphics*[width=10cm]{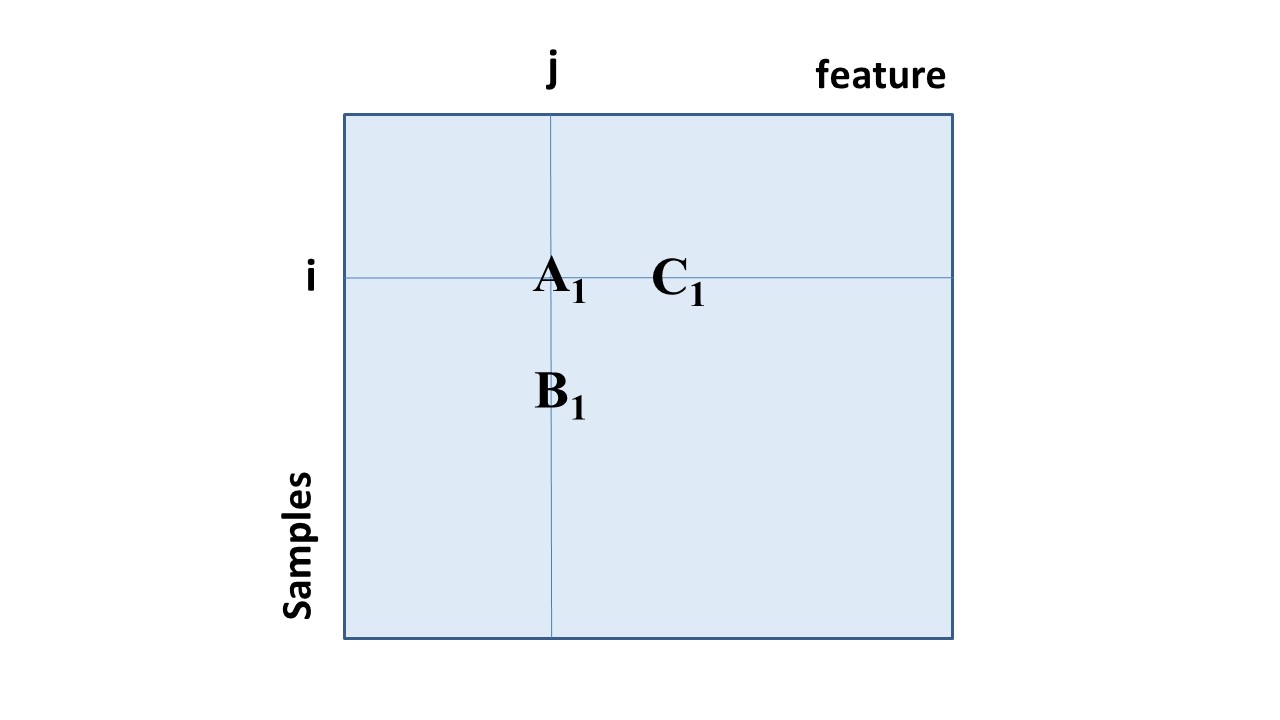}
 \caption{\footnotesize Level 1 of LC-MS measurements after Global-IS and QC correction.}
 \label{Figure7}
\end{figure}

\subsection{Level 1 measurements: alignment, QC and IS-corrected}
\label{Level 1 measurements}
One of the first steps being done after acquiring the raw data is alignment of the chromatograms, global-IS correction and QC correction of the data.  Alignment is needed to combat the lack-of-invariance problem by assuring that the same feature is now represented in a single column so that each column represents the same compound. Global-IS correction is used to reduce sample work-up and injection volume errors (see Section 9.2). The QC correction step is needed to reduce the within and between measurement batch drift of the instruments \citep{Kloet2009}.\\

\noindent After this data cleaning, we arrive at Level 1 measurements (Figure \ref{Figure7}). The columns now represent features and have the same meaning across each column. For comparing $A_1$ and $C_1$, the same argument goes as for the Level0 measurements. Comparing $A_1$ with $B_1$ is now meaningful since they pertain to the same feature and the numbers are column-conditional due to the IS and QC steps. Still, $A_1$ and $B_1$ are measured intensities and not directly interpretable as concentrations. In general, a calibration model has four regions: i) region below limit of detection (LOD), ii) the linear region, iii) a concave region (flattening) and iv) a saturated region. If $A_1$ and $B_1$ are both in region ii) then their ratio can be interpreted as a ratio of concentrations. Hence, the numbers are ratio-scaled. If they are both in the concave region iii), then a ratio is not meaningful anymore but if $A_1>B_1$ then it can still be concluded that the concentration at measurement $A_1$ is larger than the concentration at measurement $B_1$. Hence, the numbers are ordinal-scaled\footnote{Actually, a bit more than ordinal-scaled since the calibration model has a specific shape.}. If one or both of $A_1$ and $B_1$ are in the satured region iv) then a comparison is meaningless. Summarizing, the conclusion about comparability and measurement-scale in this case depends crucially on the shape of the calibration model which is unknown at this point.\\

\noindent Until now, we have been discussing \textit{numbers}. By using instrumental analysis theory into the transition from Level 0 to Level 1, we have arrived at \textit{data} because the numbers in Level 0 have become a certain meaning. In short: data=numbers+meaning.

\subsection{Level 2 measurements: group IS-corrected}
\label{Level 2 measurements}
\begin{figure}[h]
 \includegraphics*[width=10cm]{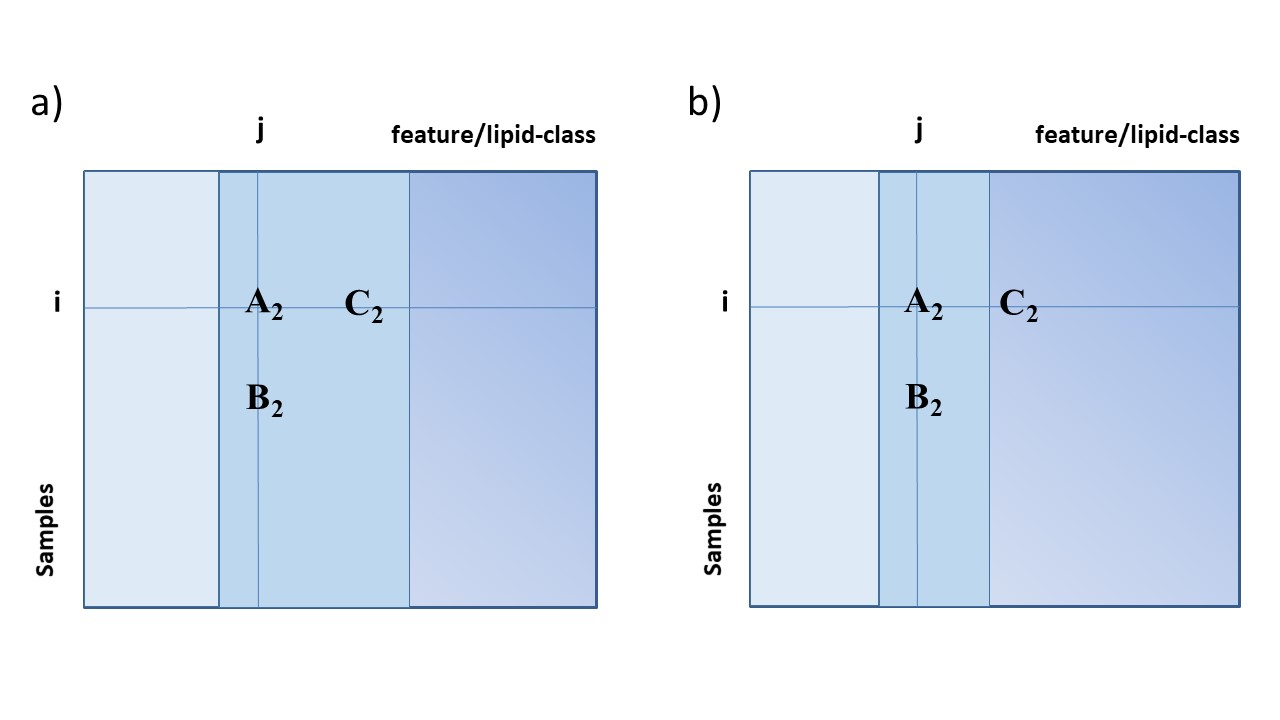}
 \caption{\footnotesize Level 2 of LC-MS measurements after Group-IS and QC correction. The panels a) and b) refer to lipid C yes/no in the same IS-class (indicated by shades of blue).}
 \label{Figure8}
\end{figure}

It is also possible to have internal standards for a group of lipids, e.g., separate standards for trigycerides and certain phospholipid classes such as phophoethanolamines, phosphoethanolserines and cholesterols. Obviously, for each rt.mz feature the lipid class has then to be assigned. Hence, at this level the metabolite or lipid class of a feature has to be identified. The results are called Level 2 measurements and shown in Figure \ref{Figure8}. For comparing $A_2$ and $B_2$ the conclusions are the same as for the Level 1 measurements. In comparing $A_2$ and $C_2$ we now have to distinguish between $A_2$ and $C_2$ in the same lipid-class (Figure \ref{Figure8}a) or not (Figure \ref{Figure8}b). When $A_2$ and $C_2$ are in the same lipid-class and if we can expect similar response factors then these numbers are comparable. Whether the numbers $A_2$ and $C_2$ are ratio- or ordinal-scaled depends again on the region of the calibration models in which $A_2$ and $C_2$ are. If the numbers $A_2$ and $C_2$ are in different lipid classes then we have in principle again Level 1 measurements. Note that in the transition from numbers to data we have not only used instrumental analysis theory but also chemical theory; in particular theory regarding the behavior during analysis (ionization) and chemical similarity between lipids.\\

\subsection{Level 3 measurements: concentrations}
\label{Level 3 measurements}
\begin{figure}[h]
 \includegraphics*[width=10cm]{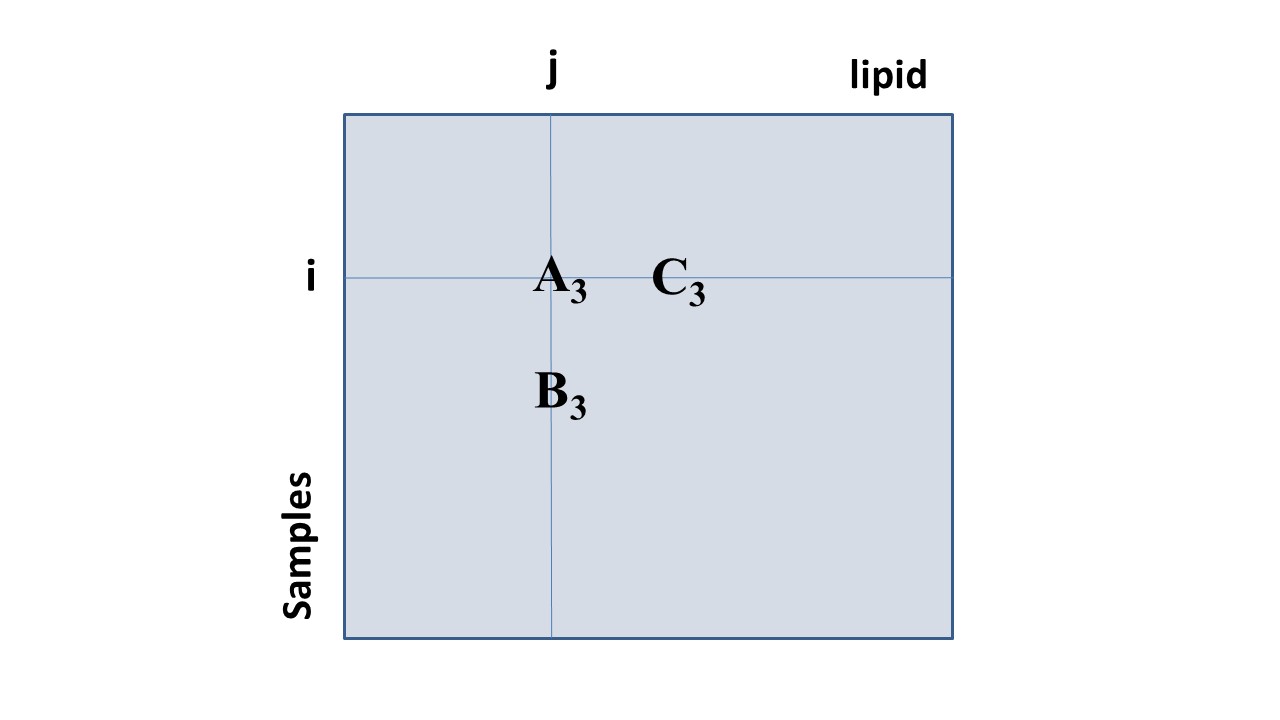}
 \caption{\footnotesize Level 3 of LC-MS measurements after using calibration models.}
 \label{Figure9}
\end{figure}

The highest level of measurements is obtained after having built calibration models for all individual lipids. Obviously, for each rt.mz feature the structure of the lipid has to be known, which is not a trivial task (but outside of the scope of this paper). The most accurate calibration models are obtained when using an authentic standard per lipid. For practical reasons, often calibration models within a class of lipids are constructed using only a limited number of standards. This is possible if the response factors are proven to be the same, or a (preferably on theory based) model for the response factor of each lipid is applied \citep{Wang2017} (see Sections \ref{Internal standards} and  \ref{Calibration models}). These allow for a transformation from intensities to concentrations for all numbers in the matrix of Figure \ref{Figure9}. This results in matrix-conditional data and the data are ratio-scaled.

\subsection{Level 4 measurements: tailor made}
\label{Biological Activities}
Up to now, the focus has been on concentrations of the lipids. Suppose that the interest is in the lipids as ligands in a biological activity study. It is known that ligands have an affinity for a receptor which can be modeled by a (nonlinear) sigmoidal dose-response function which is usually specific for each lipid. From this perspective, the data in Level 3 are now suddenly column-conditional since the values $A_3$ and $C_3$ have become incomparable due to the differences in dose-response functions. Moreover, because of the sigmoidal relationship, the values $A_3$ and $B_3$ are not ratio-scaled anymore but ordinal-scaled.\footnote{Actually, a bit more than ordinal-scaled since the dose-response curves are sigmoidal.} Hence, properties of data also depend on the biological question being asked. When all dose-response curves are known, then the data could be transformed to biological activities again and become ratio-scaled matrix conditional. This could be called Level 4 data which are tailor-made for a specific purpose.

\section{Synthesis}
\label{Synthesis}
\begin{figure}[h]
 \includegraphics*[width=10cm]{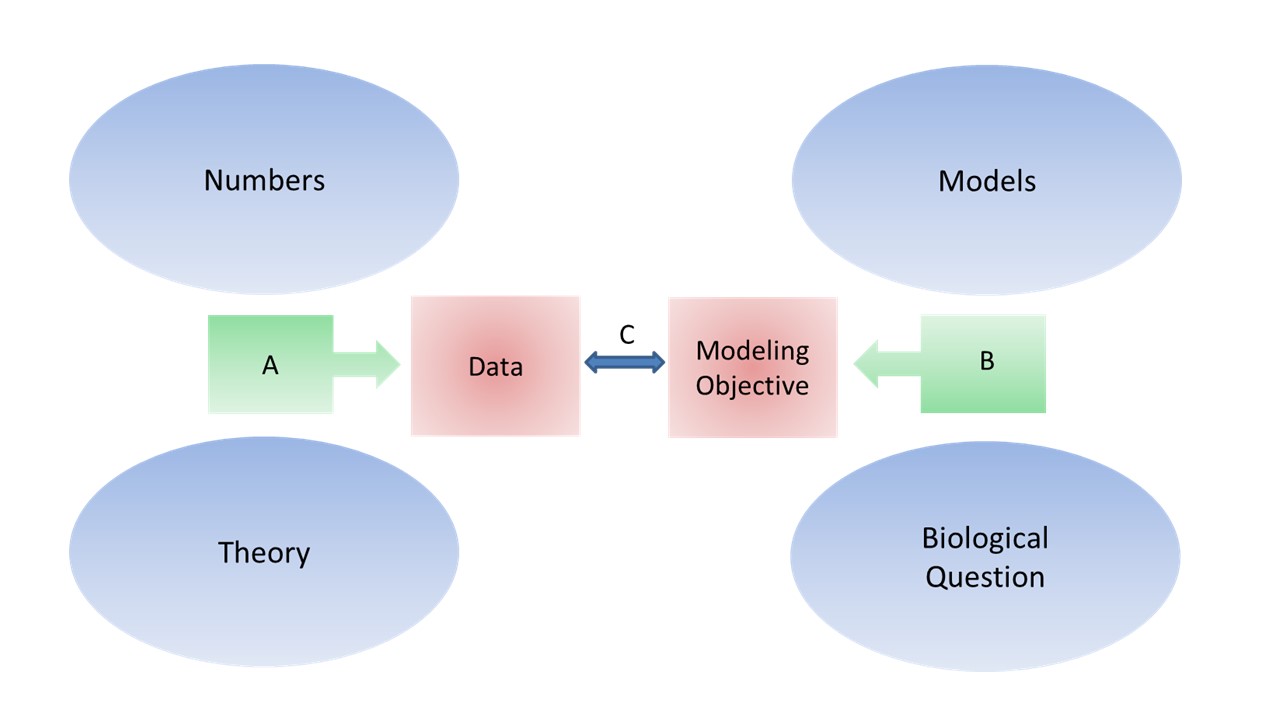}
 \caption{\footnotesize Synthesis of the foregoing. Legend: arrows A, B and C are explained in the text.}
 \label{Figure10}
\end{figure}

From the previous presentation it is clear that there is an interplay between numbers, data, theory and type of biological question. An attempt to synthesize this is shown in Figure \ref{Figure10}.\\

\noindent The blue ellipsoid marked 'Numbers' represent the raw numbers coming from an instrument. They do not represent data yet, as explained above. Instrumental analysis and chemical/physical theory should be used to turn these numbers into data (Arrow A). These data have then certain properties - conditionality, measurement scale - depending on the original numbers and the theory that is used to turn them into data. Note that this is a simplification since also the type of biological question can change the properties of the data (but see arrow C below). The biological questions pertain to certain biological systems (the ERS in terms of representational measurement theory) and these questions need to be formalized in a model to be able to confront the question with the data. The term model should be taken in a broad context, e.g., even simple correlations can be considered models. The modeling objective is then formulated in terms of which parameters have to be estimated, which loss-functions to use, which algorithms to use etc. We may, e.g, use a PARAFAC model for exploring three-way data whereby we want to study the scores and loadings or we may be using a PARAFAC model for estimating underlying parameters which are meaningful, e.g., relative concentrations in fluorescence emission-excitation spectroscopy \citep{VanMechelen2011}.\\

\noindent The crucial part of Figure \ref{Figure10} is Arrow C. There should be a match between modeling objectives and properties of the data. Citing the example above: if time-series data are not synchronized (or cannot be synchronized) then it does not make sense to use three-way models. If the data are only ordinal scaled, then we cannot fit quantitative systems biology models to the data. If there are discrepancies in Arrow C then there are two routes to take: change the properties of the data or change the modeling objective. For the example of unsynchronized time-series data we have to switch to simultaneous component analysis models (thereby possibly also rephrasing the biological question). To fit systems biology models, we have to make calibration models and make all data in concentration form. Obviously, there are many examples of how to solve such discrepancies.

\section{Broader context: considerations for the field}
\label{Broader context}

\subsection{Repercussion for metabolomics data analysis}
\label{Repercussion for metabolomics data analysis}
The above presented theory has repercussions for metabolomics data analysis. In Section \ref{Correlations} we will show what it means for correlations as an example. In Section \ref{Overview} we will subsequently give an overview.

\subsubsection{Correlations}
\label{Correlations}
To show how correlations are affected by different comparability properties, we present a small example. Suppose that intensities of three metabolites are measured at Level 1 (Global-IS and QC corrected, and aligned). The data for five samples are presented in Eqn. \ref{eIntensities}:
\beq
  \left[
  \begin{array}{ccc}
  10 & 15 & 12 \\
  12 & 10 & 8 \\
  8 & 5 & 4 \\
  10 & .. & ..\\
  6 & .. & .. \\
  \end{array}
  \right]
\label{eIntensities}
\eeq
This data is column-conditional and, depending on whether the compounds are measured within the linear range of the calibration models, ordinal or ratio-scaled within a column. The Pearson correlation matrix of this data is
\beq
  \left[
  \begin{array}{ccc}
  1 & 0.5 & 0.5 \\
  0.5 & 1 & 1 \\
  0.5 & 1 & 1 \\
  \end{array}
  \right]
\label{eIntensitiesCorr}
\eeq
assuming that the numbers are ratio-scaled. Suppose now that we have made calibration models for all three metabolites and the concentrations are as follows:
\beq
  \left[
  \begin{array}{ccc}
  5 & 3 & 6 \\
  6 & 2 & 4 \\
  4 & 1 & 2 \\
  5 & .. & ..\\
  3 & .. & .. \\
  \end{array}
  \right]
\label{eConcentrations}
\eeq
and this matrix has exactly the same (Pearson) correlations as the one of Eqn. \ref{eIntensities} (all intensities were in the linear range of the calibration models). Hence, the column-conditionality of the data does not hamper the use of correlations and when using correlations, there is no need for calibration models. The reason is that going from intensities to concentrations (assuming that the numbers are in the linear range of the calibration models) are simple linear transformations and correlations are invariant under such linear transformations. If the original intensities were on an ordinal-scale, then similarly Spearman correlations could be used.\\

\noindent Following our example, suppose that we are interested in activities, and have measured activities corresponding to the above mentioned concentrations and these are
\beq
  \left[
  \begin{array}{ccc}
  13 & 2 & 3 \\
  15 & 2 & 2 \\
  12 & 1 & 1 \\
  13 & .. & ..\\
  9 & .. & .. \\
  \end{array}
  \right]
\label{eActivities}
\eeq
where metabolite two is in the saturation phase of the dose-response curve; metabolite one also shows non-linear behavior and metabolite three is in the linear range. The correlation matrix of these activities is:
\beq
  \left[
  \begin{array}{ccc}
  1 & 0.76 & 0.33 \\
  0.76 & 1 & 0.87 \\
  0.33 & 0.87 & 1 \\
  \end{array}
  \right]
\label{eActivityCorr}
\eeq
which is clearly different from Eqn. \ref{eIntensitiesCorr} because of the non-linearity of the dose-response curves.

\subsubsection{Overview}
\label{Overview}
This section discusses the repercussions of the foregoing discussion for metabolomics data analysis. It should not be read as a cookbook about what (not) to do, but merely as some remarks about things to consider when performing metabolomics data analysis. Table \ref{Table1} summarizes the remarks.\\

\begin{table}[h!]
\begin{tabular}{llll}
\toprule
\textbf{Level} & \textbf{Characteristics} & \textbf{Data properties} & \textbf{Statistics} \\
\midrule
Level 0 & Raw numbers & Incomparable & Some within-row comparisons \\
\midrule
Level 1 & QC-corrected/aligned & Column-conditional & Within-column \\
       & Global-IS-corrected & Ordinal or ratio & Nonmetric or metric \\
\midrule
Level 2 & QC-corrected/aligned & Column-conditional & Within-column \\
       & Group-IS-corrected & Within-group matrix-conditional & Within-group submatrix \\
       &  & Ordinal or ratio & Nonmetric or metric \\
\midrule
Level 3 & Concentrations & Matrix conditional & Within matrix comparisons \\
       &                & Ratio & Metric \\
\midrule
Level 4 & Tailor made & Case specific & Case specific \\
\bottomrule
\end{tabular}
\caption{\label{Table1}\footnotesize Different levels of metabolomics measurements and their properties. For explanation, see text.}
\end{table}

\noindent The notions of conditionality and measurement scales were explained in the previous sections, and mainly summarized in Table \ref{Table1}. As also explained in the foregoing, the data obtained from Level 1 and Level 2 can be ordinal (but see the associated footnote) or ratio-scaled depending on the form of the calibration model and the specific measurement. When we are in the ordinal-scaled regime, then nonmetric methods can be applied such as Mann-Whitney two-sample tests and nonmetric multidimensional scaling \citep{Borg2005}. Also optimal scaling for multivariate analysis is then an option \citep{Gifi1990}. When we can assume ratio-scaled data then the whole machinery of PCA, PLS and OPLS-DA is at our disposal. When the data is in the ordinal-scaled regime and still methods such as PCA and OPLS(-DA) are applied, it is unclear at this point whether the results from such an analysis are (in)valid: as mentioned earlier the data is also a bit more than ordinal-scaled.

\subsection{FAIR data}
\label{FAIR}
Recently, the life sciences and especially the omics field starts to agree that data should be FAIR (Findable, Accesible, Interoperable, Reusable). This allows to re-use data, or to combine data from different sources. However, so far often not much information is provided about the quality and theory of the data: how secure is the identification of a metabolite or lipid? How quantitative are the data: is the data for a metabolites ratio-scaled or ordinal-scaled? If FAIR data is not provided with the proper measurement information and theory (i.e. meta-data), they are actually more numbers than data (see Figure \ref{Figure10}). Moreover, it has been shown in the example on biological activity that properties of data are not invariant but also depend on the biological question of interest.

\subsection{Data fusion}
\label{Data fusion}
A field of growing interest is data fusion and, specifically, fusion of metabolomics data with other types of omics data. The issues of scale-type and comparability (in general: data characteristics) also play a dominant role in this field but up to now have received little attention. An obvious question to ask is whether two data sets can be compared, likewise as comparing two columns in a matrix of measurements as explained above. When different types of omics measurements are performed on the same set of samples then such questions arise when the two data sets are going to be fused.\\

\noindent Differences in scale-type between two omics data sets also often occur, e.g., when fusing metabolomics with mutation data which are intrinsically binary. Several methods exists for fusing such types of data \citep{Mo2013,Song2018,Anderson2018,Smilde2019} but comparability issues as explained above have received little attention.

\subsection{NMR}
For NMR the situation is different than for MS-based metabolomics and we will briefly explain the levels 0-4 for NMR. At level 0, the raw NMR data is considered. These are row-conditional since values in the same column cannot be compared without preprocessing for two reasons. First, the NMR-spectra may not be aligned so that there is lack-of-invariance and, secondly, even if the spectra are aligned there may still be dilution effects (e.g. in urine spectra) hampering between sample comparisons. Within a row (that is, within the same spectrum) the numbers are comparable because they all pertain to counting of hydrogen atoms.\\

\noindent If all preprocessing has been done (aligning, calibration (e.g., ERETIC signal) and normalization) then we arrive at levels 1-2. The data are now row- and column conditional, hence, matrix conditional. In essence, the data pertains to counts of hydrogen atoms and are not concentrations yet. To arrive at concentrations, the peaks have to be identified, quantified and calibrated thus thereby arriving at level 3. These concentrations are ratio-scaled and matrix conditional. Also in this case, if interest shifts to biological activities, then the same conclusions (for level 4) hold as for the case of MS-based measurements.

\subsection{Other omics measurements}
We do not give a full treatment here but much of the theory explained above also holds for other types of omics measurements. MS-based proteomics is a clear example, but similar simple questions treated in this paper can also be asked about, e.g., RNAseq data as collected in gene-expression measurements or in microbiome research. We invite researchers in those areas to consider these simple questions to!

\section{Acknowledgements}
AS thanks Iven van Mechelen (KU Leuven, Belgium) for stimulating discussions.

\section{Appendix}

\subsection{List of abbreviations}
\label{List of abbreviations}

Table \ref{Table2} gives an overview of the abbreviations used in the main text.

\begin{table}[h!]
\begin{tabular}{ll}
\toprule
Abbreviation & Full name \\
\midrule
LC-MS & liquid-chromatography mass-spectrometry \\
GC-MS & gas-chromatography mass-spectrometry \\
CE-MS & capillary-electophoresis mass-spectrometry \\
NMR & nuclear magnetic resonance  \\
QC & quality control \\
IS & internal standard \\
PCA & principal component analysis \\
PLS-DA & partial least squares discriminant analysis \\
\bottomrule
\end{tabular}
\caption{\label{Table2}\footnotesize List of used abbrevations.}
\end{table}

\subsection{Formal treatment of measurement scales}
\label{Formal treatment of measurement scales}
A treatment of measurement scales following \citet{Stevens1946} is given below (see Table \ref{Table3}; a more rigorous explanation is given elsewhere \citep{Krantz1971,Roberts1985}) and a nice introduction is given by \citet{Hand2004}. The lowest measurement level is nominal data which are merely (exclusive) categories. Examples are different types of cars, different countries etc. The data are only used as class labels and these can be changed as long as each class receives a unique other label. Hence, the \textit{permissible transformations} - the transformations between NRS's that keep the relationships in the corresponding ERS intact - are one-to-one transformations. The type of statistics to be used for this type of data are number of cases, frequencies, $X^2$-tests, etc.\\

\begin{table}[h!]
\begin{tabular}{llll}
\toprule
Scale-type & Example & Permissible transformations & Permissible statistics \\
\midrule
Nominal & Categories & One-to-one & Number of cases \\
Ordinal & Survey data & Monotonic & Median, IQR \\
Interval & Degree Celsius & Positive linear transformation & Mean, Standard deviation \\
         & Calender time  & $x'=\alpha x+\beta\; (\alpha>0)$ & \\
Ratio & Length & Similarity transformation & Coefficient of variation \\
      & Mass   & $x'=\alpha x\; (\alpha>0)$           & \\
Absolute & Counts & $x'=x$ & All previous \\
\bottomrule
\end{tabular}
\caption{\label{Table3}\footnotesize Formal treatment of types of data scales. For explanation, see text.}
\end{table}

\noindent The next level of measurement scale are ordinal data. The prototypical example is survey data in which respondents can score on certain issues using the answers strongly disagree, disagree, neutral, agree, strongly agree. Obviously, there is an order in these answers; and these answers can be labeled from 1-5. The difference between 2 and 1 on the one hand and between 3 and 2 on the other hand - although exactly equal - does not have a meaning. The ERS can also be represented using a different set of numbers, e.g, 2,4,7,8,9 but the transformation between the two NRS's needs to be monotonic. The type of statistics to be employed are the ones for the lower-scaled measurement (i.e. nominal data) and in addition median, interquartile range (IQR) etc.\\

\noindent Interval-scale data is the next level. An example is degree Celsius where the numbers zero and hundred are arbitarily chosen. Stated otherwise, this scale does not have natural zero and unit. This means that another scale ($x'$) can be used with $x'=\alpha x+\beta\; (\alpha>0)$ and this scale has the same meaning for the ERS; an example is Fahrenheit where $\alpha=9/5$ and $\beta=32$. Nevertheless, the ratio of differences between values of this scale has meaning in terms of the ERS, e.g., in using calendar times $\frac{1980-1960}{1945-1940}=4$ can be interpreted in a meaningful way as the first period being four times as long as the second one. However, the ratio  $\frac{1980}{990}=2$ does not have a meaning; $1980$ is not 'twice as old as' $990$. Hence the name interval-scale. In addition to statistics at the lower measurement levels, also means and standard deviations can be used meaningfully for interval-scaled data.\\

\noindent The next level is ratio-scaled data with examples length and weight. As already explained in the main text, a ratio-scaled variable has no natural unit. Hence, the permissible transformation $x'=\alpha x\; (\alpha>0)$. It does have a natural zero, however, making ratio's meaningful. In addition to the lower measurement levels, also coefficients of variation can be used meaningfully for ratio-scaled data.\\

\noindent The highest degree of measurability is absolute scale data, e.g., count data. Such data has a natural zero and a natural unit and the only permissible transformation is the identity. Apart from the measurement levels mentioned above, there are still other types of more exotic scales \citet{Krantz1971}. There has been an extensive discussion about permissible statistics and measurement scales in the statistics community with opposing opinions \citep{Adams1965,Michell1986,Velleman1993,Hand1996}.

\subsection{Internal standards}
\label{Internal standards}
Internal standards are used for different purposes: i) compensate for shifts in retention time or mass calibration to support alignment of features or identification of metabolites: ii) compensate for variations in sample preparation (e.g. sample volume)  or injection volume, iii) compensate for variation in response factors due to e.g. variation in MS sensitivity and iv) support establishment of calibration models for metabolites.

The first type of correction is important for alignment of features (level 1) and for identification (prerequisite of level 2/3); the second and third type of corrections are important to allow for comparing abundances of features or metabolites between samples (level 1) or when creating semi-quantitative data (level 2) or determining concentrations (level 3). For determining concentrations often more internal standards are used, in its most extensive form one isotopically labelled internal standard per metabolite (so called isotope dilution analysis).  It should be noted that internal standards can be also used to monitor the performance of a method rather than to correct numbers to (ultimately) obtain data.

\subsection{Calibration models}
\label{Calibration models}
\begin{figure}[h!]
 \includegraphics*[width=20cm]{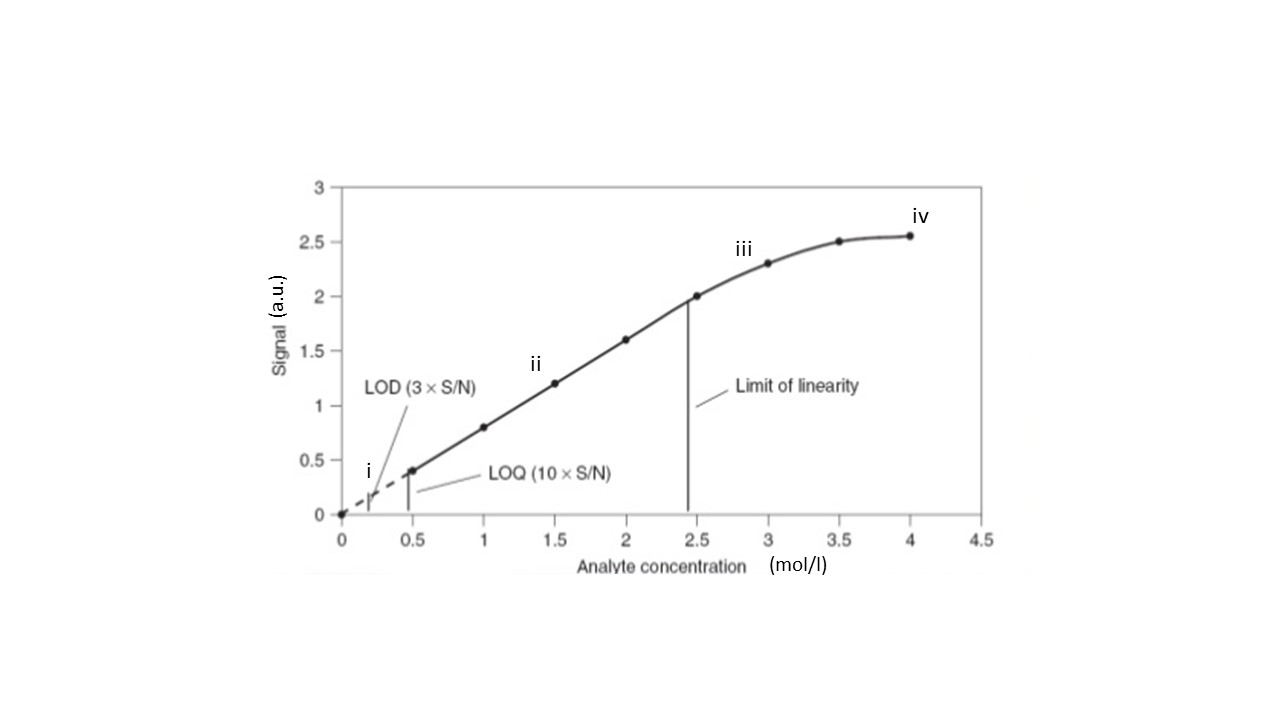}
 \caption{\footnotesize General shape of a calibration model. Abbreviations: LOD-limit is detection; LOQ is limit of quantification.}
 \label{Figure11}
\end{figure}

Calibration models (or curves) are used to make build a relation between a measured intensity (or relative intensity) and the concentration of an analyte (i.e. metabolite). In general, a calibration model consists of four regions: i) a below limit of quantification (LOQ) region, ii) a region of linearity and iii) a non-linear region and a saturation region (iv). These are indicated in Figure \ref{Figure11}.
The calibration model does not necessarily pass through the origin. It may be that the analyte is absorbed somewhere in the measurement equipment thereby generating a negative intercept or, alternatively, the analyte is present in one of the solvents as contaminant resulting in a positive intercept.\\

\noindent If only relative intensities are measured then the ratio's of such intensities do not necessarily translate to the same ratio in concentrations. This holds only for measurements in region ii. In region i), the intensities cannot be used to infer something about concentrations and in regio iii) the calibration model is nonlinear. Note that in this region, the intensities are still ordinal scaled. Actually, a bit more since the permissible transformation of ordinal-scaled variables is a monotonic increasing function, whereas in the case of calibration models, this function should also be concave.

\subsection{Concentrations}
\label{Concentrations}
An important notion in representational measurement theory is the concept of concatenation. Using the example of the lengths of the sticks this can be explained as follows. Any two sticks (of length $a$ and $b$) can be put next to each other to form one longer stick (of length $c=a+b$) and there will be a stick somewhere that has this length $c$. This process is called concatenation and obviously the ERS of sticks has this property which is reflected in the additivity property of the NRS ($c=a+b$). Concentrations do not have this property. Two solutions containing the same analyte at concentrations $a$ and $b$ do not give a solution with concentration $a+b$ when mixed. Because of the fact that two solutions with the same concentration $a$, when mixed give a solution with that concentration $a$, this measurement is called idempotent. Density is another measurement with that same property: it is also an idempotent measurement.\\

\noindent Formally, the concatenation rule for concentrations is as follows. Suppose we have two solutions having concentrations $x$ and $y$ for a certain analyte. Let $V_x$ and $V_y$ be the respective volumes, then the concentration of the solution resulting from mixing the orginal ones is
\beq
  x\circ y =\frac{xV_x+yV_y}{V_x+V_y}=(\frac{V_x}{V_x+V_y})x + (\frac{V_y}{V_x+V_y})y=\lambda x+(1-\lambda)y
\label{eConcat}
\eeq
where $\lambda \geqslant 0$ is defined implicitly and $\circ$ is the symbol for concatenation. Hence, the concentration of the mixed solution is simply the convex combination of the original ones which weights derived from their volumes. Similar rules hold for density.\\

\noindent Concentrations are derived measurements \citep{Roberts1985}. They are constructed from measuring two single measurements: moles (amounts) and volume. Concentrations are ratio-scale data; they can be measured as $mol/L$ or $mmol/L$ or $mmol/mL$ etc. Formally, moles are ratio-scaled variables\footnote{One could argue that moles are absolute-scaled since they refer to (Avogadro's) number of molecules.} and thus have the similarity transformation as their permissible transformation; likewise volume. Thus, if $x$ represents the moles, $y$ the volume and $c=x/y$ the concentration, we have
\beq
  c'=\frac{x'}{y'}=(\frac{\alpha x}{\beta y})=(\frac{\alpha}{\beta})\frac{x}{y}=\gamma \frac{x}{y}=\gamma c
\label{eConcRatio}
\eeq
where $\alpha, \beta, \gamma >0$ and $\gamma$ is implicitly defined. Thus the only permissible transformation for concentrations is the similarity one thereby defining a ratio-scale.

\subsection{pH}
\label{pH}
Another often used measurement in chemistry and metabolomics is pH. At high-school pH was defined as $-log[H^+]$, but that was later in undergraduate chemistry courses changed into $-log (a_H)$ where $a_H$ is the activity of the hydrogen ion. Both definitions point to a property of the ERS: concentrations or activity. For aqueous solutions a concept of activity is well-defined, but for non-aqueous solutions (e.g. mobile phases) this is less clear \citep{Tindall2002}. Hence, the official definition of pH from the IUPAC is now operational \citep{Buck2002}. It is defined as the reading of a standardized pH glass-electrode relative to a standard buffered solution and is an example of a measurement which cannot be described in representational theory, but instead is defined in terms of operational theory.

\bibliographystyle{natbib}
\bibliography{main}

\begin{thebibliography}{}

\bibitem[Adams {\em et~al.}(1965)]{Adams1965}
Adams, E., Fagot, R., and Robinson, R. (1965).
\newblock A theory of appropriate statistics.
\newblock {\em Psychometrika}, {\bf 30}(2), 99--127.

\bibitem[Anderson-Bergman {\em et~al.}(2018)]{Anderson2018}
Anderson-Bergman, C., Kolda, T., and Kincher-Winoto, K. (2018).
\newblock Xpca: Extending pca for a combination of discrete and continuous
  variables.
\newblock {\em arXiv preprint arXiv:1808.07510}.

\bibitem[Borg and Groenen(2005)]{Borg2005}
Borg, I. and Groenen, P. (2005).
\newblock {\em Modern multidimensional scaling}.
\newblock Springer.

\bibitem[Buck {\em et~al.}(2002)]{Buck2002}
Buck, R., Rondinini, S., Covington, A., Baucke, F., Brett, C., Camoes, M.,
  Milton, M., Mussini, T., Naumann, R., Pratt, K., Spitzer, P., and Wilson, G.
  (2002).
\newblock Measurement of ph. definition, standards, and procedures (iupac
  recommendations 2002).
\newblock {\em Pure and Applied Chemistry}, {\bf 74}(11), 2169--2200.

\bibitem[Christin {\em et~al.}(2010)]{Christin2010}
Christin, C., Hoefsloot, H. C.~J., Smilde, A.~K., Suits, F., Bischoff, R., and
  Horvatovich, P.~L. (2010).
\newblock Time alignment algorithms based on selected mass traces for complex
  lc-ms data.
\newblock {\em Journal of Proteome Research}, {\bf 9}(3), 1483--1495.

\bibitem[Coombs(1964)]{Coombs1964}
Coombs, C.~H. (1964).
\newblock {\em A theory of data}.
\newblock John Wiley \& Sons, New York.

\bibitem[Gifi(1990)]{Gifi1990}
Gifi, A. (1990).
\newblock {\em Nonlinear Multivariate Analysis}.
\newblock John Wiley \& Sons.

\bibitem[Hand(2004)]{Hand2004}
Hand, D. (2004).
\newblock {\em Measurement Theory and Practice: The World Through
  Quantification}.
\newblock John Wiley \& Sons.

\bibitem[Hand(1996)]{Hand1996}
Hand, D.~J. (1996).
\newblock Statistics and the theory of measurement.
\newblock {\em Journal of the Royal Statistical Society Series A-statistics in
  Society}, {\bf 159}, 445--473.

\bibitem[Koek {\em et~al.}(2011)]{Koek2011}
Koek, M., Jellema, R., van~der Greef, J., Tas, A., and Hankemeier, T. (2011).
\newblock Quantitative metabolomics based on gas chromatography mass
  spectrometry: status and perspectives.
\newblock {\em Metabolomics}, {\bf 7}(3), 307--328.

\bibitem[Krantz {\em et~al.}(1971)]{Krantz1971}
Krantz, D., Luce, R., Suppes, P., and Tversky, A. (1971).
\newblock {\em Foundations of Measurement (Volume I)}.
\newblock Dover.

\bibitem[Luce and Narens(1987)]{Luce1987}
Luce, R.~D. and Narens, L. (1987).
\newblock Measurement scales on the continuum.
\newblock {\em Science}, {\bf 236}(4808), 1527--1532.

\bibitem[Michell(1986)]{Michell1986}
Michell, J. (1986).
\newblock Measurement scales and statistics - a clash of paradigms.
\newblock {\em Psychological Bulletin}, {\bf 100}(3), 398--407.

\bibitem[Mo {\em et~al.}(2013)]{Mo2013}
Mo, Q.~X., Wang, S.~J., Seshan, V.~E., Olshen, A.~B., Schultz, N., Sander, C.,
  Powers, R.~S., Ladanyi, M., and Shen, R.~L. (2013).
\newblock Pattern discovery and cancer gene identification in integrated cancer
  genomic data.
\newblock {\em Proceedings of the National Academy of Sciences of the United
  States of America}, {\bf 110}(11), 4245--4250.

\bibitem[Narens and Luce(1986)]{Narens1986}
Narens, L. and Luce, R.~D. (1986).
\newblock Measurement - the theory of numerical assignments.
\newblock {\em Psychological Bulletin}, {\bf 99}(2), 166--180.

\bibitem[Roberts(1985)]{Roberts1985}
Roberts, F. (1985).
\newblock {\em Measurement theory}, volume~7 of {\em Encyclopedia of
  {M}athematics and its applications}.
\newblock Cambridge University Press.

\bibitem[Smilde {\em et~al.}(2004)]{Smilde2004}
Smilde, A.~K., Bro, R., and Geladi, P. (2004).
\newblock {\em Multiway analysis. Applications in the Chemical Sciences}.
\newblock John Wiley {\&} Sons, New York.

\bibitem[Smilde {\em et~al.}(2019)]{Smilde2019}
Smilde, A.~K., Song, Y., Westerhuis, J.~A., Kiers, H. A.~L., Aben, N., and
  Wessels, L. F.~A. (2019).
\newblock Heterofusion: Fusing genomics data of different measurement scales.
\newblock {\em arXiv}, {\bf 1904.10279}.

\bibitem[Song {\em et~al.}(2018)]{Song2018}
Song, Y., Westerhuis, J.~A., Aben, N., Wessels, L. F.~A., Groenen, P. J.~F.,
  and Smilde, A.~K. (2018).
\newblock Generalized simultaneous component analysis of binary and
  quantitative data.
\newblock {\em arXiv preprint arXiv:1807.04982}.

\bibitem[Stevens(1946)]{Stevens1946}
Stevens, S. (1946).
\newblock On the theory of scales of measurement.
\newblock {\em Science}, {\bf 103}(2684), 677--680.

\bibitem[Timmerman and Kiers(2003)]{Timmerman2003}
Timmerman, J. and Kiers, H. (2003).
\newblock Four simultaneous component models of multivariate times series from
  more than one subjcet to model intraindividual and interindividual
  differences.
\newblock {\em Psychometrika}, {\bf 86}, 105--122.

\bibitem[Tindall(2002)]{Tindall2002}
Tindall, G. (2002).
\newblock Mobile-phase buffers, {P}art {I} - the interpretation of p{H} in
  partially aqueous mobile phases.
\newblock {\em LCGC North America}, {\bf 20}(11), 1028--1032.

\bibitem[Van~der Kloet {\em et~al.}(2009)]{Kloet2009}
Van~der Kloet, F., Bobeldijk, I., Verheij, E., and Jellema, R. (2009).
\newblock Analytical error reduction using single point calibration for
  accurate and precise metabolomic phenotyping.
\newblock {\em Journal of proteome research}, {\bf 8}(11), 5132--5141.

\bibitem[Van~Mechelen and Smilde(2011)]{VanMechelen2011}
Van~Mechelen, I. and Smilde, A.~K. (2011).
\newblock Comparability problems in the analysis of multiway data.
\newblock {\em Chemometrics and Intelligent Laboratory Systems}, {\bf 106}(1),
  2--11.

\bibitem[Velleman and Wilkinson(1993)]{Velleman1993}
Velleman, P.~F. and Wilkinson, L. (1993).
\newblock Nominal, ordinal, interval, and ratio typologies are misleading.
\newblock {\em American Statistician}, {\bf 47}, 65--72.

\bibitem[Wang {\em et~al.}(2017)]{Wang2017}
Wang, M., Wang, C.~Y., and Han, X.~L. (2017).
\newblock Selection of internal standards for accurate quantification of
  complex lipid species in biological extracts by electrospray ionization mass
  spectrometry-what, how and why?
\newblock {\em Mass Spectrometry Reviews}, {\bf 36}(6), 693--714.

\bibitem[Wold {\em et~al.}(1998)]{Wold1998}
Wold, S., Kettaneh, N., Friden, H., and Holmberg, A. (1998).
\newblock Modelling and diagnostics of batch processes and analogous kinetic
  experiments.
\newblock {\em Chemometrics and intelligent laboratory systems}, {\bf 44},
  331--340.

\bibitem[Zelikovic and Chesney(1989)]{Zelikovic1989}
Zelikovic, I. and Chesney, R.~W. (1989).
\newblock Sodium-coupled amino-acid transport in renal tubule.
\newblock {\em Kidney International}, {\bf 36}(3), 351--359.

\end{thebibliography}

\end{document}